\documentclass[9pt,twocolumn,twoside]{opticajnl}
\journal{opticajournal} 

\setboolean{shortarticle}{true}

\usepackage{lineno}
\usepackage{hyperref}
\usepackage{svg}
\usepackage{subfig}
\usepackage{orcidlink}
\usepackage[utf8]{inputenc}

\title{Demonstration of a multimode-to-multimode photonic lantern for astronomy}

\author[1, *]{Marina Centenera-Merino \orcidlink{0000-0003-3332-5373}}
\author[2]{Andrew Ross-Adams\orcidlink{0009-0000-5089-8224}}
\author[2]{Christopher Betters \orcidlink{0000-0002-3797-2028}}
\author[1]{Pedro J. Amado \orcidlink{0000-0002-8388-6040}}
\author[1,3]{Jesus Aceituno \orcidlink{0000-0003-0487-1105}}
\author[4]{Kalaga Madhav \orcidlink{0000-0002-6711-7137}}
\author[3]{Stefan Cikota \orcidlink{0000-0002-7671-2317}}        
\author[3]{F. Javier Flores \orcidlink{0000-0002-2817-6038}}   
\author[4]{Julius Göhring}   
\author[4]{Abani S. Nayak \orcidlink{0000-0002-4666-9282}}   
\author[1]{Jose Luis Ortiz}   
\author[1]{David Peréz-Medialdea \orcidlink{0000-0001-5163-6429}} 
\author[1]{Francisco J. Pozuelos \orcidlink{0000-0003-1572-7707}} 
\author[4, 5]{Martin M. Roth \orcidlink{0000-0003-2451-739X}}   
\author[1]{Mabel Ruíz-López \orcidlink{0000-0002-7206-6557}}   
\author[1]{Miguel Andrés Sánchez-Carrasco \orcidlink{0000-0001-5533-3660}}   
\author[2]{Sergio Leon-Saval \orcidlink{0000-0002-5606-3874}}

\affil[1]{Instituto de Astrofísica de Andalucía, CSIC, Glorieta de la Astronomía s/n , 18008, Granada, Spain}
\affil[2]{Sydney Astrophotonic Instrumentation Laboratories, School of Physics, The University of Sydney, Australia}
\affil[3]{Centro Astronómico Hispano en Andalucía, Observatorio de Calar Alto, Sierra de los Filabres, 04550, Gérgal, Almería, Spain}
\affil[4]{Astrophotonics (innoFSPEC), Leibniz-Institut für Astrophysik Potsdam (AIP), Germany; Institut für Physik und Astronomie, Universität Potsdam, Germany}
\affil[5]{Deutsches Zentrum für Astrophysik (DZA), Postplatz 1, 02826 Görlitz, Germany}

\affil[*]{mcentenera@iaa.csic.es}

\begin{abstract}
	
Photonic lanterns have been widely used in astronomy as low-loss multiplexing devices, typically coupling light from a multimode input into several single-mode outputs. In this work, we present what we believe to be  the first multimode-to-multimode photonic lantern specifically designed to combine light from several multimode fibers into a single multimode waveguide. We fabricated and characterized the devices at multiple wavelengths to evaluate the performance of the adiabatic multimode transition. The measured efficiencies exceed $90\ \%$, demonstrating low-loss multimode propagation and efficient modal transfer through the lantern structure. This architecture enables efficient multimode beam combination and represents a significant step toward scalable modular telescope concepts without requiring diffraction-limited injection.

\end{abstract}

\setboolean{displaycopyright}{false} 

\begin{document}
	
	\maketitle
	
	\section{Introduction}
	\label{sec:Introduction}

    Large-aperture telescopes enable frontier science but involve significant manufacturing, operational, and financial challenges \citep{Liske2012}. Modular telescope architectures offer a scalable alternative, provided that light from multiple sub-apertures can be efficiently combined. Projects such as Photonic EMARCOT \citep{centenera2025marcot} explore this approach, employing multimode photonic lanterns (MM-PLs) as key coupling elements.

    Photonic lanterns (PLs) are adiabatic waveguide transitions that enable low-loss coupling between multimode (MM) and single-mode (SM) systems \citep{leon2005multimode, birks2015photonic}. Conventionally, they operate as multimode-to-single-mode converters, preserving étendue through mode-number matching.		

    These devices have been widely used in astronomy due to their high efficiency. In this context, they are typically employed as splitters to divide the light coming from a MM fiber into several SM fibers, providing a more stable output. Their ability to enhance light collection efficiency has also been demonstrated in free-space coupling applications \citep{ozdur2013free, tedder2024application}. To date, and to our knowledge, only one group \citep{moraitis2023opa} has explored PLs operating as a combiner for astronomical applications; however, that work employs conventional lanterns. In seeing-limited conditions, the turbulence of the atmosphere distorts the incoming wavefront, producing a speckled intensity pattern that cannot be efficiently coupled into a SM fiber. Achieving high coupling efficiency in this regime typically requires adaptive optics systems (AO), which are complex and costly.

    However, the extension of this concept to multimode-to-multimode coupling has remained largely unexplored. The aim of this work is to fabricate a novel PL configuration operating as a combiner, in which light propagates from several step-index MM fibers into a single MM fiber. The use of larger fiber cores enables a greater portion of the sky to be collected, relaxing the need for costly AO elements under seeing-limited conditions, making it well suited for astronomical and free-space optical applications.
    
    Unlike fiber bundle combiners, which suffer from fill-factor losses due to interstitial cladding material and do not conserve étendue at the junction, the MM-PL forms a single monolithic waveguide through adiabatic fusion tapering, enabling coherent supermode evolution and efficient modal transfer.

    The telescope MARCOT-Pathfinder \citep{roth2022marcot}, is currently operating at the Calar Alto Observatory (CAHA, Spain) and will serve to validate the overall concept and assess the on-sky performance of the MM-PL. A key technological challenge in this context is the efficient combination of multiple seeing-limited multimode beams into a single output compatible with high-resolution spectroscopic instrumentation.
    
	\section{Materials and Methods}
	\label{sec:Materials_Methods}

    The MM-PLs presented in this work were designed to operate as beam combiners for the MARCOT-Pathfinder telescope. The device geometry and optical parameters were selected to preserve étendue conservation while efficiently coupling several multimode telescope feeds into a single multimode output waveguide. Table \ref{tab:Materials_Pars} summarizes the most relevant parameters of fabricated MM-PLs.
    
	
	These materials were selected based on the design requirements of the telescope, which consists of seven individual optical tube assemblies. Accordingly, seven MM input fibers were selected. The core diameter of these fibres was selected to collect at least twice the median seeing disk at CAHA ($\sim 0.9$ arcsec \cite{sanchez2007night}). Given that the input beam has a focal ratio of f/5 and the effective plate scale of the telescope is $\rm{101.61\ "/mm}$, by choosing $25\ \mu m$ core diameter fibers, we can inject  $\rm{2.54}$ arcsec.

	For a configuration consisting of seven input fibers, each with a core diameter of $25\ \mu$m and a numerical aperture of $NA = 0.1$ ($\equiv f/5$), the next step is to determine the number of spatial modes supported by each fiber. This is given by  the V-number (Eq. \ref{eq:V_M}), where $a$ is the fiber radius, $NA$ is the numerical aperture, and $\lambda$ is the wavelength, from which one can estimate the number of guided modes $M$ using
    
    \begin{equation}
        \begin{aligned}
            V &= \frac{2\pi a}{\lambda}NA, \quad
            M \approx \left(\frac{V}{2}\right)^2
        \end{aligned}
    \label{eq:V_M}
    \end{equation}

    Calculating $M$ is essential for ensuring étendue conservation, defined by the optical invariant $A \cdot \Omega = \text{const}$, where $A$ is the beam cross-sectional area and $\Omega$ is the angular extent of the beam; which requires mode-number matching between both ends of the MM-PL. As the fibers are tapered together, increasing modal overlap between neighboring fibers causes the guided modes of the individual cores to progressively couple and evolve into collective supermodes \citep{xia2011supermodes} supported by the combined structure.
    
    A critical requirement is that the taper transition be adiabatic, evolving sufficiently slowly to suppress inter-modal coupling and prevent mode leakage \citep{birks2015photonic}. A smooth asymmetric taper profile was used ($\rm{550\ mm}$ taper-down, $\rm{20\ mm}$ waist), as detailed in \href{https://opticapublishing.figshare.com/articles/journal_contribution/Supplementary_document_for_First_demonstration_of_a_multimode-to-multimode_photonic_lantern_for_astronomy_-_8011966_pdf/33198921?file=68151580}{Supplement 1}, and the high transmission efficiencies observed suggest that the transition approaches the adiabatic regime.
    
    For the parameters considered in this work, the modal content can be explicitly estimated. At a wavelength of $\rm{520\ nm}$, each input fiber supports approximately $V \approx 15.1$, corresponding to $M \approx 57$ guided modes per fiber. This yields a total of $\sim 399$ modes at the input of the lantern (7-port PL). For the output waveguide, the C1-based lantern (NA = 0.196, see Table \ref{tab:Materials_Pars}) supports $V \approx 59.2$, corresponding to $M \approx 876$ modes, while the C2-based lantern (NA = 0.132) supports $V \approx 39.9$, corresponding to $M \approx 397$ modes. These values confirm that the C1-based lantern comfortably satisfies the étendue conservation condition ($M_{out} \approx 876 \gg M_{in} \approx 400$). The C2-based lantern, however, operates at the design limit ($M_{out} \approx 397 \approx M_{in}$), where any slight deviation in fabrication parameters could result in a marginal mode mismatch. This can be straightforwardly addressed in future fabrication runs by slightly increasing the output core diameter, which would provide the necessary modal margin while preserving the overall lantern design.
	

    The numerical aperture of the photonic lantern output, $NA_{PL}$, is determined by the capillary used during fabrication and is given by $NA_{PL} = \sqrt{(NA_{cap})^2 - (NA_{fi})^2}$, where $NA_{cap}$ and $NA_{fi}$ are the numerical apertures of the capillary and the input fibers, respectively. The resulting $NA_{PL}$ values for each lantern design are listed in Table \ref{tab:Materials_Pars}.

    Under the condition of mode-number matching between all $N_{fi}$ input fibers and the output waveguide, the required output core radius can be estimated using Eq. (\ref{eq:PL_radius}), where $N_{fi}$ is the number of input fibers, and $a_{fi}$ and $NA_{fi}$ are the radius and numerical aperture of the input fibers.
	
    \begin{equation}
        \label{eq:PL_radius}
        a_{PL} = \frac{\sqrt{N_{fi}} \cdot a_{fi} \cdot NA_{fi}}{NA_{PL}}
    \end{equation}

Applying this relation yields minimum output core diameters of $33.75\ \mu$m for lanterns fabricated with C1 capillaries, and $49.59\ \mu$m for those using C2 capillaries. To ensure that all guided modes are supported across both designs, an output core diameter of $50\ \mu$m was selected.

	\begin{table}
		\centering
		\begin{tabular}{ccc}
			\toprule
            \toprule
            \textbf{Capillary} & \textbf{ID ($\mu m$)} & \textbf{NA} \\
            \midrule
            C1 & $750$ & $0.220$ \\
            C2 & $415$ & $0.166$ \\
            \midrule
			\textbf{Photonic Lantern} &  \textbf{Core diameter ($\mu m$)}&  \textbf{NA}\\
			\midrule
			PL 1, PL 2 & $50$ & $0.196$ \\
			PL 3, PL 4 & $50$ & $0.132$ \\ 
            \midrule
            \textbf{Fibers} & \textbf{Core diameter ($\mu m$)}&  \textbf{NA}\\
            \midrule
            Bare & $25$ & $0.100$ \\
            FC/PC-terminated & $25$ & $0.130$ \\
			\bottomrule
		\end{tabular}
		\caption{Estimated output waveguide parameters of the fabricated MM-PLs. PL 1 and PL 2 correspond to photonic lanterns manufactured using C1 capillaries (C1-based lanterns), whereas that PL 3 and PL 4 correspond to use C2 capillaries (C2-based lanterns).}
		\label{tab:Materials_Pars}
	\end{table} 
	
    The MM-PLs used in this study were fabricated using a 3SAE Combiner Manufacturing System (CMS) from step-index MM fibers inserted into fluorine-doped capillaries. The fabrication process includes fiber preparation, loading, and tapering to form the multimode-to-multimode transition. For further details of the fabrication process, see \href{https://opticapublishing.figshare.com/articles/journal_contribution/Supplementary_document_for_First_demonstration_of_a_multimode-to-multimode_photonic_lantern_for_astronomy_-_8011966_pdf/33198921?file=68151580}{Supplement 1}.

    After fabrication, the efficiency of each lantern was evaluated. The bare fibers of the MM-PL were spliced to standard connectorized MM fibers ($\rm{NA = 0.13}$, FC/PC-terminated) to enable measurement. For this purpose, both the lantern fibers and the standard fibers were stripped, cleaved, and subsequently joined by fusion splicing.
    
    All efficiency measurements reported in this work were performed on bare MM-PLs. Two independent characterization methods were employed to verify the measured efficiencies and to assess the contribution of splice and insertion losses. 

    The method described in the following paragraphs inherently accounts for splice-related losses, since the reference measurement includes the spliced region; therefore, possible insertion losses introduced by the splice were taken into account.

    For the insertion method (\href{https://opticapublishing.figshare.com/articles/dataset/Insertion_losses_measurements_at_543nm_and_white_source_csv/32789814}{Data File 1}) described in the following paragraphs, additional splice and coupling losses were independently estimated and corrected. Further details can be found in \href{https://opticapublishing.figshare.com/articles/journal_contribution/Supplementary_document_for_First_demonstration_of_a_multimode-to-multimode_photonic_lantern_for_astronomy_-_8011966_pdf/33198921?file=68151580}{Supplement 1}.
	
	A tunable laser source operating at two wavelengths, $\rm{520\ nm}$ and $\rm{685\ nm}$, was used. The FC/PC fiber was connected to the source, while the output of the photonic lantern was measured by a power meter.
		
	The maximum and minimum power values detected by the photodiode at the MM-PL output were recorded, and their average value was calculated. This measurement was repeated three times to obtain a median value. For the reference measurement, the fibers were cleaved after the splice region, and the power at the output of these cleaved fibers was measured. By applying the cut-off method, the potential losses introduced by the splices were taken into account.

We applied this second measurement method to revalidate the results obtained previously. The procedure was very similar to the previous setup. In this case, the light sources used were a HeNe laser operating at 543 nm and a continuous white light source. This method was applied to PL 4 (see Table \ref{tab:Materials_Pars}).

In this second case, all seven ports were spliced to standard fibers and illuminated individually, obtaining an output power value for each fiber. The efficiency was then calculated by dividing the measured output power by the reference power.

The reference power was determined using two identical multimode fibers connected end-to-end, and the transmitted power was recorded. The reference measurement was repeated at the start and end of the process, and the mean value was used for the efficiency calculation.

\section{Results and Discussion}
\label{sec:Results}

The fabricated MM-PLs exhibit high transmission efficiency across all tested input ports, with median efficiencies above $90\ \%$ for the measured wavelengths. The fiber numbering is device-specific and does not reflect a consistent spatial mapping across lanterns; details are provided in \href{https://opticapublishing.figshare.com/articles/journal_contribution/Supplementary_document_for_First_demonstration_of_a_multimode-to-multimode_photonic_lantern_for_astronomy_-_8011966_pdf/33198921?file=68151580}{Supplement 1}.

\begin{figure*}
	\centering
	\subfloat{
		\includegraphics[width=0.45\textwidth]{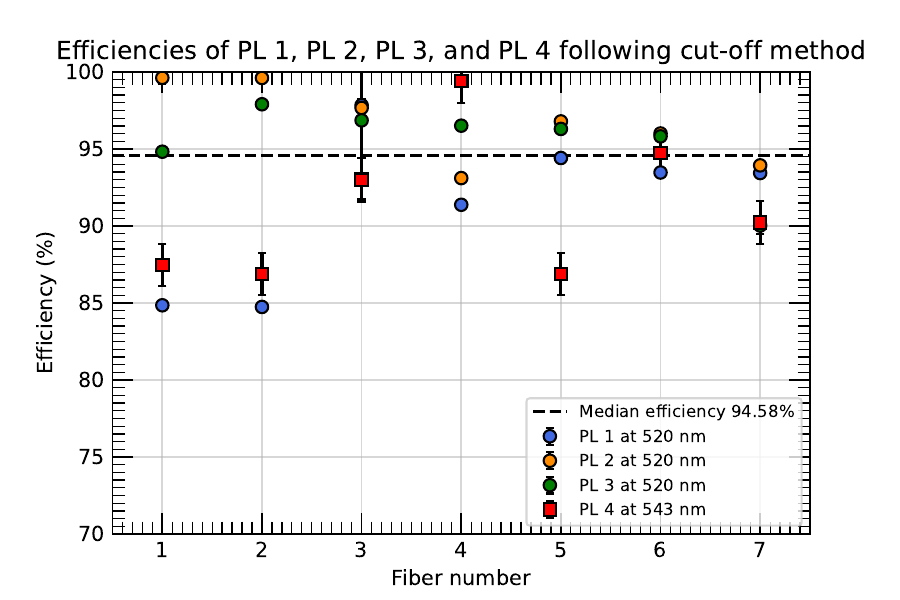}}
	\subfloat{
		\includegraphics[width=0.45\textwidth]{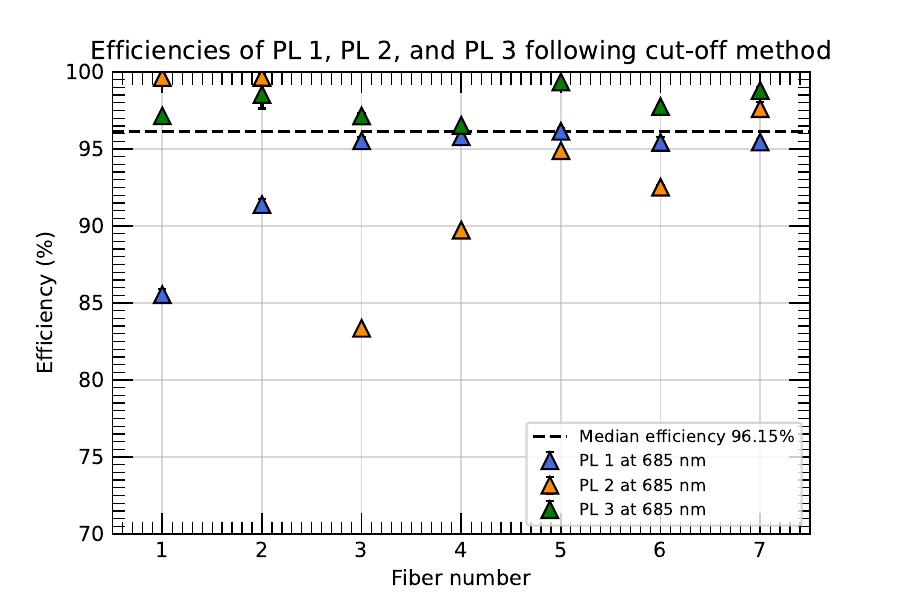}}
	\caption{All devices exhibit efficiencies above $\rm{80\ \%}$, with median values of $\rm{94.58\ \%}$ at $\rm{520\ nm}$ (left) and $\rm{96.15\ \%}$ at $\rm{685\ nm}$ (right). The colors of the data points correspond to measurements from different MM-PLs, where the same color at both wavelengths corresponds to the same MM-PL. See \href{https://opticapublishing.figshare.com/articles/dataset/PL1_measurements_cut_off_method_at_520nm_and_685nm_csv/32789829?file=65857083}{Data File 2}, \href{https://opticapublishing.figshare.com/articles/dataset/PL2_measurements_cut_off_method_at_520nm_and_685nm_csv/32789838?file=65858199}{Data File 3}, \href{https://opticapublishing.figshare.com/articles/dataset/PL3_measurements_cut_off_method_at_520nm_and_685nm_csv/32789844?file=65858328}{Data File 4}, and \href{https://opticapublishing.figshare.com/articles/dataset/PL4_measurements_cut_off_method_at_543nm_csv/32789847?file=65858331}{Data File 5} for underlying values}
	\label{fig:efficiency_no_packed_SAIL}
\end{figure*}

\begin{figure}
	\centering
	\includegraphics[width=.9\linewidth]{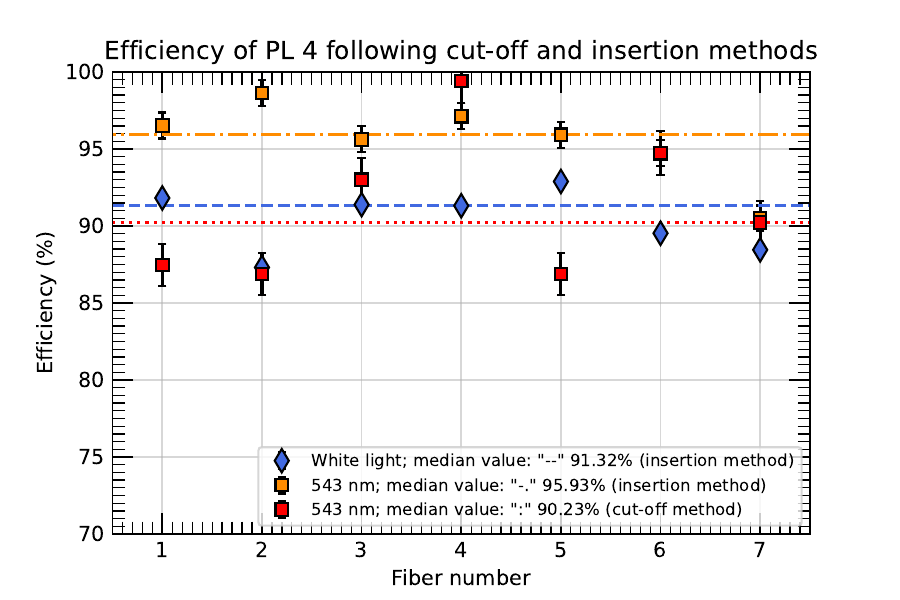}
	\caption{Efficiencies of PL 4 measured using both methods. The median efficiencies obtained were $\rm{95.93\ \%}$ at $\rm{543\ nm}$ and $\rm{91.35\ \%}$ for the white light source following insertion method, meanwhile the median was $\rm{90.23\ \%}$ at $\rm{543\ nm}$ following cut-off method. See \href{https://opticapublishing.figshare.com/articles/dataset/PL4_measurements_insertion_method_at_543nm_and_white_source_csv/32789850?file=65858334}{Data File 6} for underlying values.}
	\label{fig:Eff_Insertion_method}
\end{figure}

Our MM-PL efficiencies, measured at $\rm{520, \ 543, \ 685\ nm}$, as well as under broadband illumination, exceed $90\ \%$ for all fabricated devices and for both characterization methods (see Figs. \ref{fig:efficiency_no_packed_SAIL} and \ref{fig:Eff_Insertion_method}). In the case of the insertion method, these values have been corrected for additional losses introduced by fiber-to-fiber coupling, as detailed in \href{https://opticapublishing.figshare.com/articles/journal_contribution/Supplementary_document_for_First_demonstration_of_a_multimode-to-multimode_photonic_lantern_for_astronomy_-_8011966_pdf/33198921?file=68151580}{Supplement 1}.

The two values below 90\% correspond to ports 1 and 2 of PL 1, 
the first device fabricated, where process parameters were not yet 
fully optimized. The slightly larger uncertainties of PL 4 at 
$\rm{543\ nm}$ reflect the lower stability of the HeNe source relative to 
the tunable laser used at $\rm{520}$ and $\rm{685\ nm}$.

A tendency toward slightly lower efficiencies at shorter wavelengths is observed in several devices. This is consistent with the increased number of supported modes at shorter wavelengths (Eq. (\ref{eq:V_M})), which enhances modal mismatch at the lantern transition. Furthermore, higher-order modes, which are more numerous at shorter wavelengths, are more susceptible to leakage during the taper, as they are less tightly confined and therefore more sensitive to geometric imperfections in the waveguide structure. Additional contributions from splice quality, fiber alignment, and modal dispersion are also likely to play a role in the observed wavelength dependence.

Each measurement was repeated three times; reported values are medians with standard deviation error bars.

The insertion-method efficiencies are systematically lower than those obtained with the cut-off method, this could be due to the fact that the pigtail fibers used for connectorization ($\rm{NA = 0.13}$) have a slightly higher numerical aperture than the input fibers ($\rm{NA = 0.1}$). This mismatch over-fills the lantern acceptance cone, introducing higher-order modal content, and may introduce an additional coupling loss at the splice.

\begin{figure}
	\centering
	\includegraphics[width=.9\linewidth]{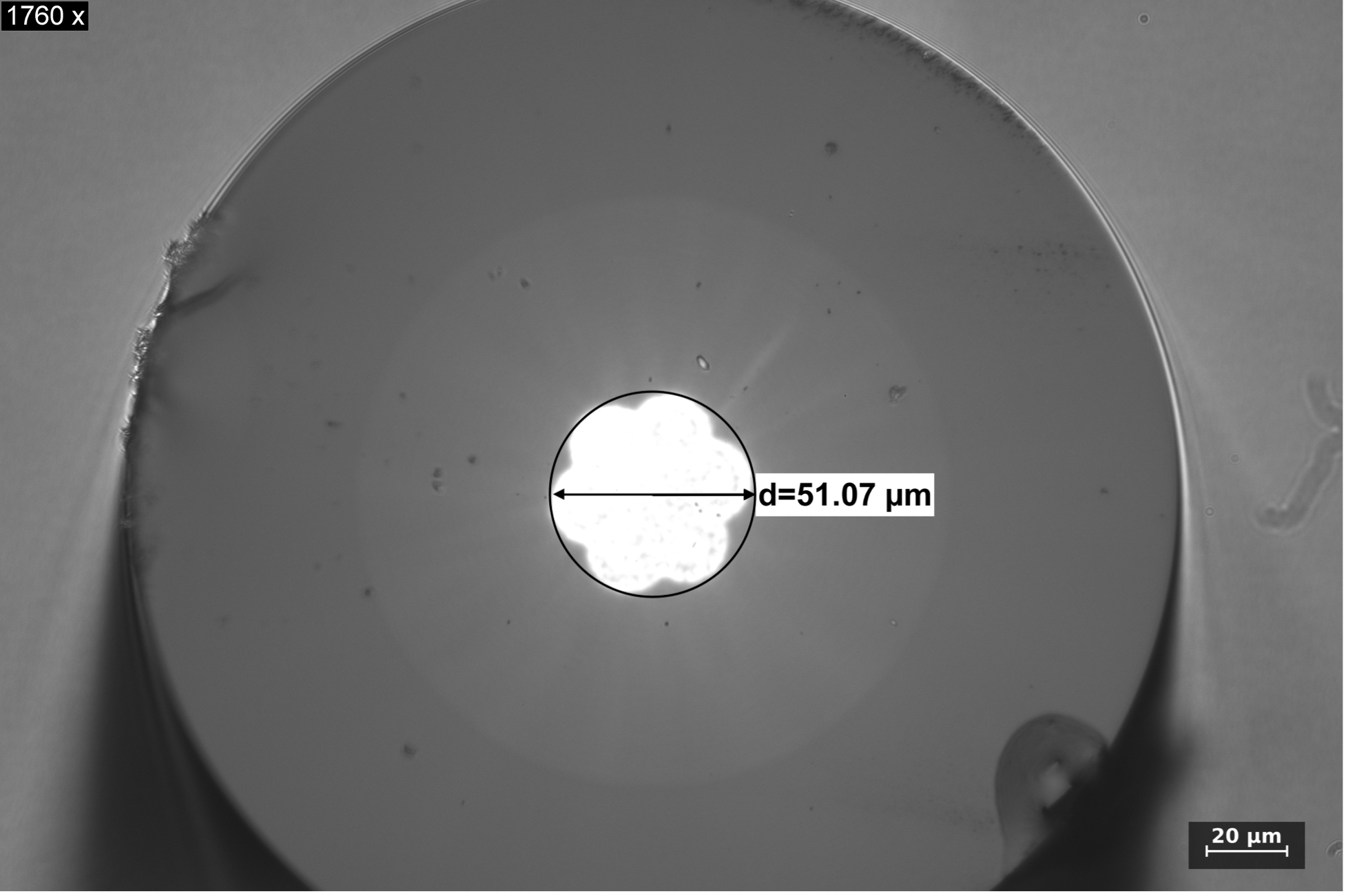}
	\caption{Microscope end-face image of the lantern output using C2 capillary (PL 4). The measured core diameter is $51.07\ \mu m$ ($40.47\ \mu m$ core at the minimum, no F-doped material included).}
	\label{fig:PL_tube_2_microscope}
\end{figure}

The near-field image at the output end (Fig. \ref{fig:PL_tube_2_microscope}), obtained by illuminating one of the input fibers with white light, shows how the optical power remains confined within the new guiding region formed by the seven fused fibers, resulting in a characteristic flower-shaped structure. This behavior confirms that the tapered structure operates as a single MM waveguide rather than as a set of independent fiber cores.

Together, these measurements confirm that the fabricated devices operate as efficient MM-to-MM combiners with low insertion loss and reproducible behavior across the seven input channels, demonstrating their potential for scalable modular telescope architectures.

\section{Conclusions}
\label{sec:Conclusions}

We have demonstrated a high-efficiency multimode-to-multimode photonic lantern capable of combining light from seven multimode input fibers into a single multimode output waveguide. The fabricated devices achieve efficiencies above $\rm{90\ \%}$ across the tested wavelengths, confirming low-loss multimode propagation and effective modal transfer through the tapered transition.

Comparable efficiencies were obtained for devices fabricated with different capillaries, indicating that étendue conservation, rather than the specific capillary type, is the dominant requirement for high intrinsic transmission. Although the insertion method yields slightly lower efficiencies due to additional coupling losses, both methods show consistent trends and comparable values. This agreement within experimental uncertainties validates the reliability of the measurements.

Port-to-port uniformity is good across all devices, with remaining variations attributable to fabrication tolerances.

Future work will focus on improving reproducibility, developing robust packaging, and scaling the number of input fibers. These results establish MM-to-MM photonic lanterns as viable high-efficiency coupling elements for scalable modular telescope architectures.



\begin{backmatter}
\bmsection{Funding} This thesis is carried out with the funding of a contract from the State Research Agency (AEI) aimed at predoctoral contracts for the training of research personnel (FPI). Specifically, the scholarship corresponds to the Severo Ochoa project of the IAA SEV-2017-0709-21-2 in the 2022 call. Likewise, the contract is part of the research activities of the CEX2021-001131-S-20-9 project and the one requested in early 2024 for the following three years. Also to the i-Link project ("MARCOT-Pathfinder: Science and Technology Development"), supporting international collaboration with AIP (Germany) and SAIL (Australia).

\bmsection{Acknowledgment}  This work was funded by the Spanish National Research Council (CSIC) through the i-LINK 2024 programme, project ILINK24064. We acknowledge financial support from the Agencia Estatal de Investigación (AEI/10.13039/501100011033) of the MICIU and the ERDF “A way of making Europe” through projects PID2022-137241NB-C43, from the Centre of Excellence “Severo Ochoa” award to the Instituto de Astrofísica de Andalucía (CEX2021-001131-S), and from the project AST22-00001-8 of the Junta de Andalucía and the Ministerio de Ciencia, Innovación y Universidades funded by the NextGenerationEU and the Plan de Recuperación, Transformación y Resiliencia. Also we acknowledge project C17.I01.P02.S17.SI01 Andalucía-Astrofísica $\&$ Física de Altas Energías.

\bmsection{Disclosures} The authors declare no conflicts of interest.

\bmsection{Data Availability Statement}  Data underlying the results presented in this paper are available in \href{https://opticapublishing.figshare.com/articles/dataset/Insertion_losses_measurements_at_543nm_and_white_source_csv/32789814?file=65856876}{Data File 1} \cite{datasetcentenera2026}.

\bmsection{Supplemental document}
See \href{https://opticapublishing.figshare.com/articles/journal_contribution/Supplementary_document_for_First_demonstration_of_a_multimode-to-multimode_photonic_lantern_for_astronomy_-_8011966_pdf/33198921?file=68151580}{Supplement 1} for supporting content.
\end{backmatter}




\bibliography{biblio}





\end{document}